\documentclass[11pt,preprint]{aastex}

\slugcomment{To appear in ApJ}

\shorttitle{Limits on the event rates of fast radio transients}
\shortauthors{Wayth et al.}

\begin{document}


\title{Limits on the event rates of fast radio transients from the V-FASTR experiment}


\author{Randall B. Wayth and Steven J. Tingay}
\affil{International Centre for Radio Astronomy Research, Curtin University. GPO Box U1987, Perth WA, 6845. Australia}
\email{randall.wayth@icrar.org}

\author{Adam T. Deller}
\affil{ASTRON, Oude Hoogeveensedijk 4, 7991 PD Dwingeloo, The Netherlands}

\author{Walter F. Brisken}
\affil{NRAO, PO Box O, Socorro, NM 87801, USA}

\author{David R. Thompson, Kiri L. Wagstaff and Walid A. Majid}
\affil{Jet Propulsion Laboratory, California Institute of Technology, 4800 Oak Grove Drive, Pasadena, CA 91109, USA}

\begin{abstract}
We present the first results from the V-FASTR experiment, a commensal search for fast transient radio bursts using the Very Long Baseline Array (VLBA).
V-FASTR is unique in that the widely spaced VLBA antennas provide a discriminant against non-astronomical signals and a mechanism for the localization and identification of events that is not possible with single dishes or short baseline interferometers.  Thus far V-FASTR has accumulated over 1300 hours of observation time with the VLBA, between 90 cm and 3 mm wavelength (327 MHz - 86 GHz), providing the first limits on fast transient event rates at high radio frequencies ($>$1.4~GHz).  V-FASTR has blindly detected bright individual pulses from seven known pulsars but has not detected any single-pulse events that would indicate high redshift impulsive bursts of radio emission.  At 1.4~GHz, V-FASTR puts limits on fast transient event rates comparable with the PALFA survey at the Arecibo telescope, but generally at lower sensitivities, and comparable to the ``fly's eye" survey at the Allen Telescope Array, but with less sky coverage.  We also illustrate the likely performance of the Phase 1 SKA dish array for an incoherent fast transient search fashioned on V-FASTR.
\end{abstract}


\section{Introduction}
High time resolution probes of the Universe at radio wavelengths, historically focused on the study of pulsars, are increasingly being employed in the search for short duration ($<1$~s) single pulses of radio emission (``fast transients'') from explosive events in the nearby or distant Universe.  A series of recent intriguing observations of apparently highly dispersed fast transients \citep{2007Sci...318..777L, 2011MNRAS.415.3065K, ban11} suggest an astronomical origin for these events, but the use of single dish radio telescopes precluded their precise localization on the sky, and the pursuit of the underlying physical mechanisms driving them.
Meanwhile, events having similar characteristics reported by \citet{2011ApJ...727...18B}, suggest a non-astronomical origin.
Again, the limitations of single dish radio telescopes make these events difficult to interpret.
Explosive events visible at extragalactic distances require extreme physical conditions; \citet{2010PASA...27..272M}, \citet{2012ApJ...744..109S}, and references therein provide excellent summaries of known and possible exotic source populations. Furthermore, measuring the dispersion of the impulsive radio emission would directly measure the column of ionized material between the event and Earth.  If a host galaxy can be unambiguously identified, this can be separated into contributions from the intergalactic medium and the interstellar media in our Galaxy and the host galaxy.  Hence, observations of fast radio transients could provide one of the few possible probes of the ionized intergalactic medium, believed to comprise more than half the baryons in the local universe \citep{2008Sci...319...55N}.

As a method of unambiguously detecting fast transients and locating them on the sky, \citet{2011ApJ...735...97W} and \citet{2011ApJ...735...98T} recognise the power of interferometric arrays of widely spaced radio telescopes, which are robust to local radio frequency interference posing as astronomical radio emission.  The utility of interferometric telescopes for the detection and localization of this class of event has prompted the development of new methods of detecting fast transients using interferometers \citep{2012ApJ...744..109S,2011ApJS..196...16B,2011arXiv1112.0308L}. Interferometer-based experiments are planned or underway at LOFAR \citep{2011A&A...530A..80S}, the GMRT \citep{2011BASI...39..353B}, ASKAP \citep{2010PASA...27..272M}, the MWA \citep{2009IEEEP..97.1497L} and the ATA \citep{2012ApJ...744..109S,2011ApJ...742...12L}.  

Experiments such as V-FASTR, along with theoretical analyses \citep{2011ApJ...734...20M}, are important to help inform the design of experiments with the next generation of wide-field radio telescopes, eventually leading to experiments using the SKA.  In this Letter, we present the results of the first $\sim$12 months of V-FASTR data collection from commensal analysis of regular VLBA observations.

\section{Observations}
The V-FASTR experiment and initial trial observational results have previously been described by \cite{2011ApJ...735...97W} and \cite{2011ApJ...735...98T}; the reader is referred to these papers for a detailed description.  V-FASTR has been undertaken under VLBA proposals BT100, BT111, BT118 (P.I. Tingay) and BM348 (P.I. Majid).

VLBA data are correlated at the Array Operations Center in Socorro, NM. During correlation, spectrometer data are generated with time resolution between 1~ms and 2~ms. In the detection stage of the system the dedispersed time series are time averaged over two, four and eight time steps to provide maximum sensitivity to pulses with intrinsic widths between approximately 1~ms and 10~ms. Based on previous evidence from \citet{2007Sci...318..777L} and \citet{2011MNRAS.415.3065K}, 10~ms was deemed an appropriate upper value for the averaging time, especially at frequencies above 1.4 GHz where the effects of scatter broadening decrease dramatically.  Trial dispersion measures between 0 and 5000~$\mathrm{cm}^{-3}$~$\mathrm{pc}$ are used. The VLBA's total observing bandwidth can optionally be split into several non-contiguous sub-bands, so the spacing between trial dispersion measures (DMs) is tailored to each observation's maximum and minimum frequency and integration time.

About half of all VLBA observations employ standard continuum observing modes which are optimal for V-FASTR purposes.  There are several other VLBA observing modes that the V-FASTR pipeline can process with full effectiveness, as briefly described below, making up the vast majority of the remaining half of the observing time.  One class of observation can make use of two receivers simultaneously.  Specifically, the 2.4~GHz and 8.4~GHz receivers are used as a pair in geodetic observations; the V-FASTR pipeline does not discriminate based on the dual-band configuration and will completely search DM-space just as for any other observation, treating the split band as a single band.  Second, pulsar experiments often employ a ``gate'' to disable accumulation of cross correlations during the off-phases of the pulsar. In the DiFX correlator \citep{2007PASP..119..318D,2011PASP..123..275D}, the auto correlations are not affected by the gate and thus from the perspective of V-FASTR, the data look the same as continuum data.  Finally, some observations target spectral lines, some of which are strong enough to affect the observed band-pass. By design the V-FASTR system automatically subtracts stable spectral features before dedispersion is performed, so the presence of spectral lines does not affect performance.


\section{System performance}
Some pulsars trigger the detector and serve as good tests of the system.
V-FASTR blindly detected pulsars J0157$+$6212, J0332$+$5434, J0826$+$2637, J1136$+$1551, J1607$-$0032, J1919$+$0021 and J1935$+$1616 during normal operations at the correct DMs. It is worth noting that pulsars J0157$+$6212, J0826$+$2637, and J1919$+$0021 have mean peak flux densities of 140~mJy, 500~mJy, and 50~mJy, respectively \citep{1995MNRAS.273..411L,2005AJ....129.1993M}, which is below our detection threshold. In these cases, a small number of individual pulses substantially exceeded the average flux density and were caught by the system. The NE2001 model of Galactic electron density distribution \citep{2002astro.ph..7156C} predicts high modulation at 1.4 GHz due to diffractive interstellar scintillation for J0826$+$2637 and J0157$+$6212, but not for J1919$+$0021.  For J0157$+$6212 and J1919$+$0021, the effects of interstellar scintillation are insufficient to push the intensity of individual pulses above the V--FASTR detection threshold -- pulse-to-pulse variability is required.  Individual pulsars show substantial differences in single--pulse behaviour \citep{2012MNRAS.tmp.2900B} and neither J0157$+$6212 nor J1919$+$0021 have been previously studied in this manner.  However, neither the detection of J0157$+$6212 nor J1919$+$0021 implies variability at a level inconsistent with the population analysis of \citet{2012MNRAS.tmp.2900B}.

We have used these pulsar detections to tune our thresholds for the triggering and coherent follow-up of candidate events.  To date, V-FASTR has amassed thousands of detections of individual pulses with detection signal-to-noise ratio (SNR) of 5 or greater.  Using captured baseband data, we can reliably obtain antenna delay solutions for events with a detection SNR as low as 8. Using these delay solutions, we can recover the position of the transient source at the level of several arcseconds without external calibration data \citep{2011ApJ...735...97W}. Delay solutions can sometimes be obtained from lower SNR events, but success is not guaranteed. Based on this experience, we set the SNR threshold for follow-up to be 7$\sigma$, at which level we typically receive of order 1-10 candidate events per day.  So this threshold is also logistically feasible for manual candidate inspection and classification, as described in \citet{2011ApJ...735...97W} and \citet{2011ApJ...735...98T}. 

Radio frequency interference (RFI) at one or more stations can be a significant practical bottleneck to survey sensitivity. V-FASTR uses adaptive noise reduction to accommodate its diverse system configurations and observing frequencies. A multi-station detection strategy \citep{2011ApJ...735...98T} distinguishes local interference from true astronomical events; it models separate stations as independent measurements of a common signal, with an independent interference process at each station producing sporadic additive noise events.  A robust statistical estimator excises one or more stations' extreme values at each new time step; such strategies have been shown to improve the effective sensitivity for a fixed budget of false triggers.

The characteristics of RFI at the VLBA vary across the spectrum and are somewhat different in nature at the different VLBA sites\footnote{http://www.vlba.nrao.edu/astro/rfi/}. Most continuum observations make use of standard frequency setups that avoid the worst RFI. Generally speaking, the interference is worse at lower frequencies where antenna directivity is lower, stray signal scattering is more effective, and transmitters are stronger and more numerous. Fortunately, most intermittent RFI can be recognized by the lack of dispersion sweep. RFI above 20~GHz is very rarely seen. Between 4 and 20~GHz, intermittent narrow-band RFI is occasionally seen, tending to be local to individual sites. Most RFI encountered at the VLBA is seen in the 20~cm and 13~cm bands where constellations of satellites (e.g., GPS, Glonass, Iridium, and satellite radio) transmit. It is in these bands that the signal classification methods employed in this work are both most effective and most necessary.

Satellites or other local RFI can coincidentally appear at more than one station simultaneously, and under these conditions optimal sensitivity requires removal of multiple streams per timestep.  However, the intrinsic sensitivity is reduced as more extreme values are adopted, so determining the appropriate excision level for each new observation is tantamount to a one-parameter optimization.  We set this value on-line during the detection process by periodically injecting synthetic pulses into the datastream just before the incoherent de-dispersion step.  These pulses are injected at known intervals and dispersion measures with SNRs ranging from 5 to 9.  A self tuning system tests excision levels between 0 and 4 stations, using the value which results in the best retrieval rate for injected pulses before the first false candidate trigger.  We re-estimate the optimal excision level every 10000 timesteps during the scan to handle time-varying noise.

Following the automated steps described above, the resultant candidate events each day are manually inspected to determine the likelihood of an astronomical origin.  If deemed astronomically plausible via inspection, the data can then be reprocessed and imaged as described above as final confirmation.



\section{Results and discussion}
\label{sec:discussion}
Other than the detections of known pulsars, V-FASTR has not yet detected any single radio pulses.  Table \ref{tab:mainresults} summarizes the number of hours of data processed by the V-FASTR system, grouped by VLBA receiver band (restricted to VLBA observations with 64 MHz bandwidth), including only observations that had five or more VLBA antennas. Table \ref{tab:mainresults} lists the time-weighted average number of antennas that participated in the observations at each frequency, used in the calculation of event rate limits below (e.g. at 2 cm, sensitivity is calculated for an array of 9.5 antennas, rather than 10).

\begin{table*}
  \caption{V-FASTR sensitivity and event rate limits for VLBA observing bands, as of 2012 May.}
  \centering
  \begin{tabular}{ r | r r rc c c c}
        & Freq  & FWHM      &N$_{ant}\tablenotemark{a}$& SEFD          & $7\sigma $    & Time  & Event rate \tablenotemark{b} \\
Receiver & (MHz) & (arcmin) && (Jy)          & sens (Jy)     & (hr)  & limit (deg$^{-2}$hr$^{-1}$) \\ \hline
    90cm & 325  & 160       &9.6& $\sim$2200    &   5         & 7.2   & 0.03 \\
    50cm & 610  & 83        &NA& $\sim$2200    &   5         & 0  & NA \\ 
    20cm & 1500 & 34        &9.4& $\sim$300     &   0.7        & 318   & 0.01 \\ 
    13cm\tablenotemark{c} & 2200 & 23  &9.4& $\sim$300 & 0.7   & 112   & 0.08 \\ 
    6cm & 4800  & 11        &9.8& $\sim$200     &   0.5        & 14    & 2.89 \\ 
    4cm & 8400  & 6         &10.0& $\sim$300     &   0.7        & 217   & 0.59 \\ 
    2cm & 14000 & 4         &9.5& $\sim$600     &   1         & 185   & 1.92 \\ 
    1cm & 22000 & 2         &9.8& $\sim$500     &   1         & 301   & 2.90 \\ 
    7mm & 43000 & 1         &9.4& $\sim$1400    &   3         & 163   & 20.5 \\ 
    3mm\tablenotemark{d} & 86000 & 0.6       &7.7& $\sim$4000    &   10         & 18    & 747 \\ 
\end{tabular}
\tablenotetext{a}{Time-weighted average of the number of antennas used for V-FASTR observations.}
\tablenotetext{b}{Calculated for the primary beam main lobe only assuming a top-hat response over the FWHM.}
\tablenotetext{c}{This row includes dual 2.4/8.4~GHz geodetic observations}
\tablenotetext{d}{Presently only 8 antennas have a 3mm receiver, which is accounted for in this row.}
\label{tab:mainresults}
\end{table*}

Following the treatment of fast transient event rates for the PALFA survey \citep{2009ApJ...703.2259D}, we calculate similar limits derived from the V-FASTR data.  PALFA used a seven beam receiver on Arecibo, whereas V-FASTR uses the ten distributed antennas of the VLBA, leading to some modification of the event rate and sensitivity calculations.  A significant difference between the calculations is that \citet{2009ApJ...703.2259D} ignore the contribution to the event rate provided by the near sidelobes of the Arecibo antenna.  For the work presented here, we have included the event rate contribution due to the first and second sidelobes of the VLBA antennas (assuming circularly symmetric sidelobes), with 0~dB far sidelobes the sole contributor to the event rate limit beyond the 3$^{rd}$ null of the antenna beam.  The inclusion of the contributions due to the first and second sidelobes has the effect of significantly improving the event rate limit at sensitivities two orders of magnitude worse than the boresight sensitivity, by virtue of the large solid angle subtended by those sidelobes.

\begin{figure}
 \centering
 \includegraphics[angle=270, scale=0.5]{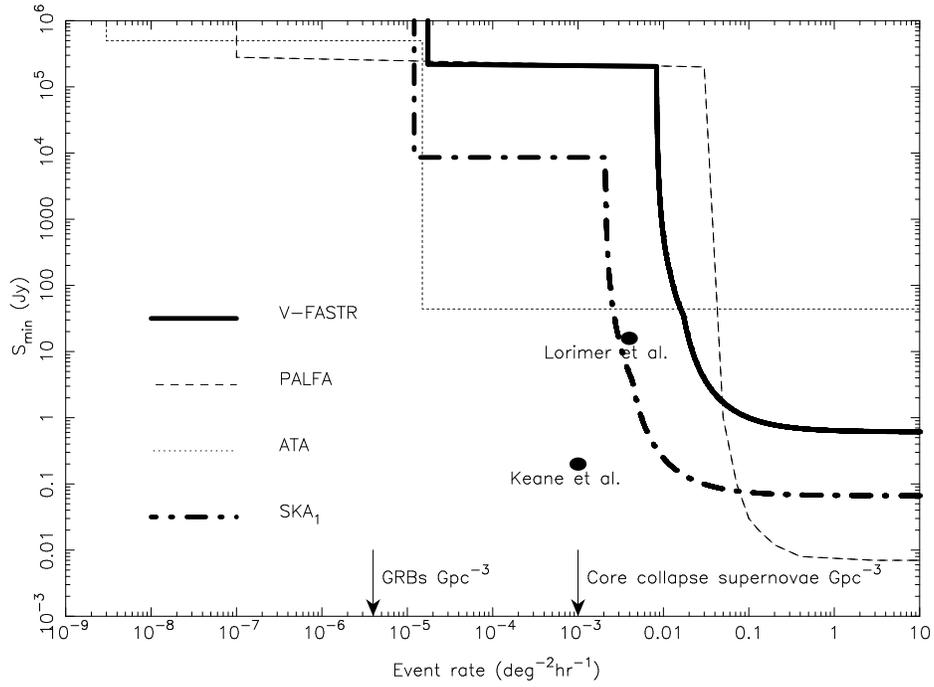}
  \vspace{5mm}
 \caption{V-FASTR event rate limits (solid line) at 1.4 GHz compared to limits also at 1.4 GHz from \citet{2009ApJ...703.2259D} (dashed line) and \citet{2012ApJ...744..109S} (dotted line).  The event rates inferred from the Lorimer and Keane bursts are shown, as are the event rates for GRBs and core collapse supernovae used by \citet{2012ApJ...744..109S}.  Also shown are the event rate limits for the Phase 1 SKA dish array, as described in the text (dot-dashed line).}
 \label{fig:limits_20cm}
\end{figure}

Figure \ref{fig:limits_20cm} shows the V-FASTR event rate limits relative to the PALFA limits, based on the data presented in Table \ref{tab:mainresults} for the V-FASTR 1.4~GHz observations.  In addition to the information in Table \ref{tab:mainresults}, the following characteristics of the VLBA antennas have been used in the limit calculations: fraction of hemisphere covered by far sidelobes = 0.5; antenna efficiency = 0.7.  At event flux densities greater than 600 mJy (10~ms integration), the PALFA and V-FASTR limits on event rate are very similar.  

While the PALFA survey is an example of a narrow/deep survey, an example of a wide/shallow survey is the Allen Telescope Array (ATA) ``fly's eye" survey \citep{2012ApJ...744..109S}, which covered 150 deg$^{2}$ of sky over 450 hr at 1.4 GHz and with a sensitivity of 44 Jy (at the same averaging period of 10~ms).  Figure \ref{fig:limits_20cm} also shows the limits arising from the ATA experiment, which do reach the order of event rate implied by \citet{2007Sci...318..777L} and \citet{2011MNRAS.415.3065K}.

We follow Figure 11 of \citet{2012ApJ...744..109S} and indicate a number of fiducial points on Figure \ref{fig:limits_20cm}, including the rates infered from the suggested extragalatic events reported by \citet{2007Sci...318..777L} and \citet{2011MNRAS.415.3065K}, as well as the same event rates (Gpc$^{-3}$) for Gamma-Ray Bursts and core collapse supernovae as cited by \citet{2012ApJ...744..109S}, to provide context for the V-FASTR/PALFA/``fly's eye" comparison.

While our limits at 1.4 GHz have not yet reached the fiducial points shown in Figure \ref{fig:limits_20cm}, V-FASTR observations are accumulated in commensal fashion and the limits will change in two ways.  First, because of the pending high bandwidth upgrade to the VLBA, the V-FASTR limit curve will move to higher sensitivities (move down in Figure \ref{fig:limits_20cm}), by approximately a factor of three.  Second, as more time is spent on sky (currently at a rate of 100 - 200 hr/month at 1.4 GHz), the limit curve will move to exclude less frequent events (to the left in Figure \ref{fig:limits_20cm}).  From Figure \ref{fig:limits_20cm}, it can be seen that with the sensitivity upgrade and a factor of $\sim$10 increase in observing time, V-FASTR will be reaching an interesting region of parameter space in terms of the events reported by \citet{2007Sci...318..777L} and \citet{2011MNRAS.415.3065K} at 1.4 GHz.

\begin{figure}
 \centering
 \includegraphics[angle=270,scale=0.5]{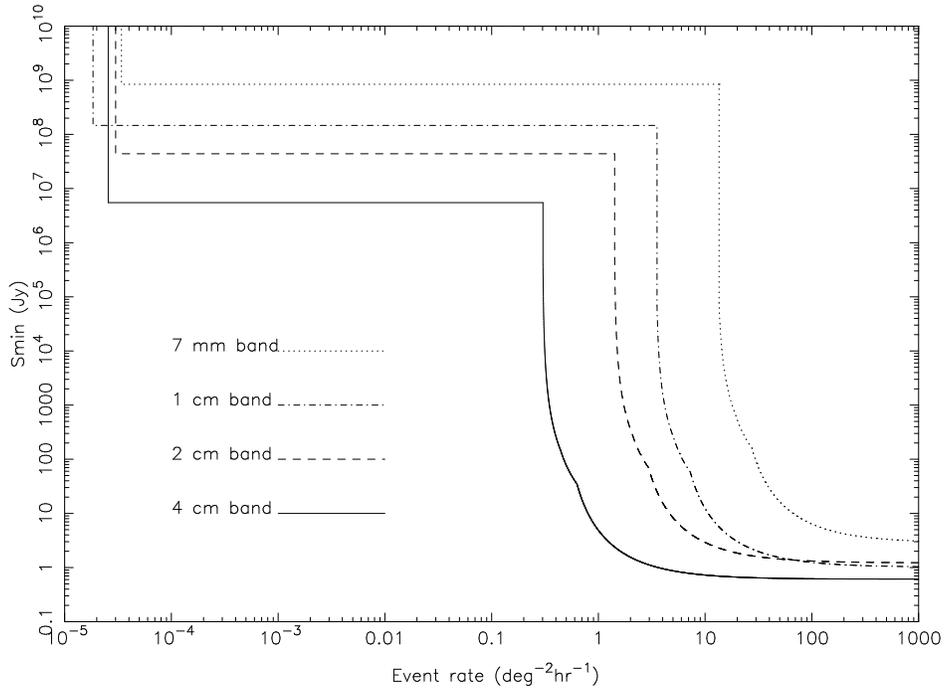}
 \vspace{5mm}
 \caption{V-FASTR event rate limits for the 4~cm, 2~cm, 7~mm, and 3~mm bands}
 \label{fig:limits_others}
\end{figure}

However, Figure \ref{fig:limits_20cm} only reflects the V-FASTR data at 1.4~GHz, representing 24\% of the total V-FASTR dataset.  One significant strength of the V-FASTR experiment is that the VLBA operates at frequencies above 1.4~GHz and we can thus derive the first limits on fast transient event rates at frequencies higher than $\sim$1.4~GHz.  Figure \ref{fig:limits_others} shows limits derived in the same manner as above for V-FASTR data at 4~cm, 2~cm, 1~cm and 7~mm, representing 16\%, 14\%, 23\% and 12\% of the total V-FASTR dataset, respectively.  The remaining percentage of time is dominated by the first observations made using the new VLBA wideband system, which will be reported separately in a future publication.

\begin{figure}
 \centering
 \includegraphics[scale=0.5]{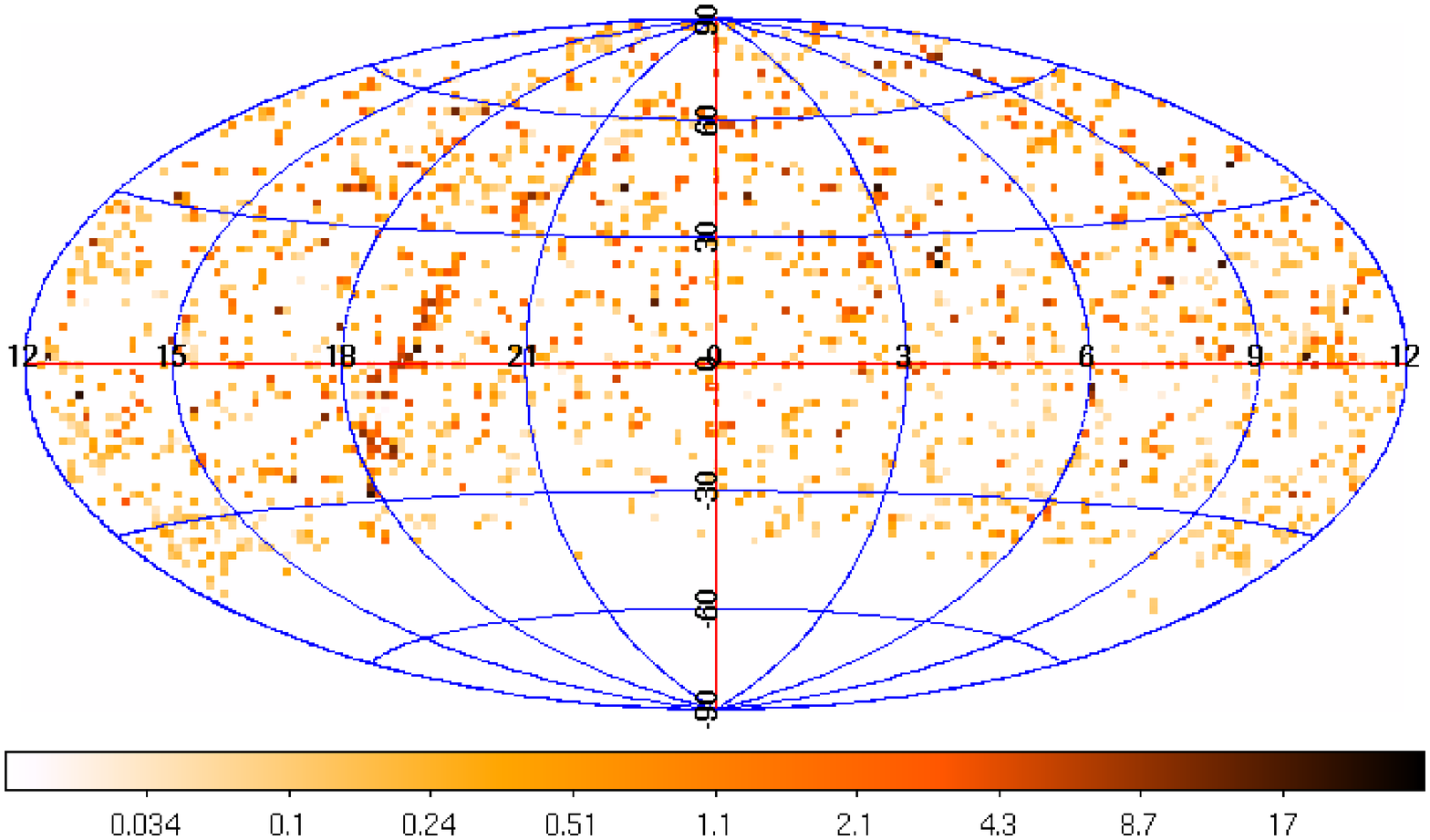}
 \vspace{5mm}
 \caption{Combined time observed per square degree of sky over all VLBA receivers (colorbar indicates observation time in hours, logarithmic scale). The sky is shown in equatorial coordinates.  Pixels in this figure span 4 deg$^{2}$.  Note the overdensity of observations following the Galactic plane.}
 \label{fig:sky_cov_all}
\end{figure}

Our event rate limits assume an isotropically distributed source population, in which case any telescope pointing direction is as likely to see an event as any other. In practise, where the VLBA is pointed depends strongly on the observing frequency. Observations at 20~cm often point at pulsars so are biased towards the Galactic plane, whereas 4~cm observations mostly target quasars.
Figure \ref{fig:sky_cov_all} shows the sky coverage of the V-FASTR observations, combined over all frequencies.  Observations are roughly equally distributed over the entire sky visible to the VLBA, plus an over-density of observations along the Galactic plane. Pulse broadening due to ionized material in the Galactic plane could potentially bias our results. To estimate the magnitude of this effect, we used the NE2001 Galactic free electron density model \citep{2002astro.ph..7156C} to calculate the expected pulse broadening of each observation, assuming the source is at a distance of 1~Mpc. In only 2\% of our observations would an extragalactic pulse have been scatter-broadened more than 1~ms, so we conclude that this is not a significant source of error, given that we are sensitive to pulses up to 10~ms duration.

Given the paucity of information for the fast transient events thus far detected with single dish experiments \citep{2007Sci...318..777L,2011ApJ...727...18B,2011MNRAS.415.3065K} due to very limited angular resolution, the fact that they have all been detected with a single instrument (Parkes radio telescope), and all in a single observing band (20~cm), obtaining a diversity of search parameters is very important for future investigations of fast transients.  In particular, without knowledge of the spectral index distribution of fast transients, searches across a wide range in frequency, such as provided by V-FASTR for the first time, are required.

While the existence of fast radio transients as a signature of high energy explosive events at cosmological distances is uncertain, the science return from detecting and localizing even a small number of such astronomical events is extremely high, as a unique and direct probe of the intergalactic medium.  V-FASTR is a low cost and highly efficient method of probing the poorly explored fast transient parameter space and is a trailblazer for what is possible in the future with new instruments under development, in particular the SKA and its Precursors and Pathfinders.  These large and sensitive interferometers, if designed such that high time and frequency resolution autocorrelation data can be accessed in parallel to the regular signal path in real time, will be able to support V-FASTR style experiments.  

With the high sensitivities and very wide fields of view planned for the next-generation instruments, V-FASTR style experiments will make rapid in-roads into the relevant parameter spaces.  To illustrate this point, Figure \ref{fig:limits_20cm} shows the expected results of a commensal fast transient survey between 1 and 2 GHz with the Phase 1 SKA (SKA$_{1}$) dish component, using the SKA$_{1}$ specifications listed in \citet{Dewdney2010} and the same survey duration as PALFA, 461 hr.  After $\sim$4000 hours, the event rate limit provided by SKA$_{1}$ would cover the combined rate and luminosity implied by \citet{2007Sci...318..777L} and \citet{2011MNRAS.415.3065K} and will retain the crucial localization characteristic of the V-FASTR experiment, as a highly distributed array, although not with VLBA-scale antenna spacings (SKA$_{1}$ will be limited to $\sim$100~km baselines).  

\acknowledgements
The International Centre for Radio Astronomy Research is a Joint Venture between Curtin University and The University of Western Australia, funded by the State Government of Western Australia and the Joint Venture partners. S.J.T is a Western Australian Premier’s Research Fellow.  R.B.W is supported via the Western Australian Centre of Excellence in Radio Astronomy Science and Engineering. A.T.D was supported by an NRAO Jansky Fellowship and an NWO Veni Fellowship. Part of this research was carried out at the Jet Propulsion Laboratory, California Institute of Technology, under contract with the US National Aeronautics and Space Administration. The National Radio Astronomy Observatory is a facility of the National Science Foundation operated under cooperative agreement by Associated Universities, Inc.  This research has made use of NASA's Astrophysics Data System.  We thank the anonymous referee for very useful comments that improved the paper.

{\it Facility:} \facility{VLBA}.


\end{document}